\documentstyle[12pt,amsfonts]{article}
\newcommand{\svec}[1]{ \stackrel{\rightarrow}{#1} }

\newcommand{\tdot}[1]{ \stackrel{\cdot}{#1} }

\newcommand{\define}{ \stackrel{\triangle}{=} }

\def\be{\begin{equation}}
\def\ee{\end{equation}}
\def\ba{\begin{array}}
\def\ea{\end{array}}

\def\d4{{\rm d}^4}

\begin{document}
%---------------------------------------------------
\vskip -1.0cm
\title{\bf Spin-Spin Interactions in Gauge Theory of Gravity, Violation
            of Weak Equivalence Principle and New Classical Test of
            General Relativity}
\author{ {Ning WU}\thanks{email address: wuning@mail.ihep.ac.cn}
\\
\\
{\small Institute of High Energy Physics, P.O.Box 918-1,
Beijing 100049, P.R.China}
}
\maketitle
%\vskip 0.8in
%\noindent

%\vskip 0.8in
%\noindent

\begin{abstract}
For a long time, it is generally believed that spin-spin interactions
can only exist in a theory where Lorentz symmetry is gauged, and
a theory with spin-spin interactions is not perturbatively renormalizable.
But this is not true. By studying the motion of a spinning particle in gravitational
field, it is found that there exist spin-spin interactions in gauge theory
of gravity. Its mechanism is that a spinning particle will generate
gravitomagnetic field in space-time, and this gravitomagnetic field will
interact with the spin of another particle, which will cause spin-spin
interactions. So, spin-spin interactions are transmitted by gravitational
field. The form of spin-spin interactions in post Newtonian approximations
is deduced. This result can also be deduced from the Papapetrou equation.
This kind of interaction will not affect the renormalizability
of the theory. The spin-spin interactions will violate the weak equivalence
principle, and the violation effects are detectable. An experiment is
proposed to detect the effects of the violation of the weak equivalence
principle.

\end{abstract}


%-------------------------------------------------------

\newpage

\Roman{section}

\section{Introduction}

It is know that the source of gravitational interactions is energy-momentum.
In Newton's classical theory of gravity\cite{01}, the motion of a test
particle in gravitational field is driven by the gravitational force on
the particle, and the spin of the particle will not affect its motion
in gravity. In general relativity\cite{02,03}, the motion of a test
particle in gravitational field is determined by geodesic equation.
But the motion of a spinning particle is not given by geodesic equation.
In general relativity, the motion of a spinning particle is given
by the Papapetrou equation, which is deduced from the Bianchi identities
and under the assumption of pole-dipole approximation\cite{04,05,06,07}.
\\

Quantum Gauge Theory of Gravity(QGTG) is first proposed in
2001\cite{08,09,10,11}. It is a quantum theory of gravity
proposed in the framework of quantum gauge field theory.
In 2003,  Quantum Gauge General Relativity(QGGR)
is proposed in the framework of QGTG\cite{12,13,14}.
Unlike Einstein's general theory of relativity, the cornerstone
of QGGR is the gauge principle, not the principle of equivalence,
which will cause far-reaching influence to the theory of gravity.
In QGGR, the field equation of gravitational gauge field is just
the Einstein's field equation, so in classical level, we can set up
its geometrical formulation\cite{15}, and QGGR returns to
Einstein's general relativity in classical level.  The
field equation of gravitational gauge field in QGGR is the same as
Einstein's field equation in general relativity, so two equations
have the same solutions, though mathematical expressions of the
two equations are completely different.
For classical tests of gravity, QGGR gives out the same theoretical
predictions as those of GR\cite{16}, and for non-relativistic
problems, QGGR can return to Newton's classical theory
of gravity\cite{17}. Based  on the coupling
between the spin of a particle and gravitoelectromagnetic
field, the equation of motion of spin can be obtained in QGGR.
In post Newtonian approximations, this equation of motion of
spin gives out the same results as those of GR\cite{18}.
QGGR is a perturbatively
renormalizable quantum theory, and based on it, quantum effects of
gravity\cite{19,20,21,22} and gravitational interactions of some basic
quantum fields \cite{23,24} can be explored. Unification of fundamental
interactions including gravity can be fulfilled in a semi-direct
product gauge group\cite{25,26,27,28}. If we use the mass
generation mechanism which is proposed
in literature \cite{29,30}, we can propose a new
theory on gravity which contains massive graviton and
the introduction of massive graviton does not affect
the strict local gravitational gauge symmetry of the
action and does not affect the traditional long-range
gravitational force\cite{31}. The existence of massive graviton
will help us to understand the possible origin of dark matter.
\\

The equivalence principle is generally believed to be one of the milestone
of Einstein's  general relativity. It has been tested by many
experiments\cite{31a,31b,31c,31d,31e}. All these experiments give out
null result for the violation of equivalence principle. On the other hand,
it is generally believed that spin-spin interaction between rotating
bodies will violate equivalence principle. Zhang et al. proposed a
phenomenological model for spin-spin interactions\cite{31f}. Later, Zhang and
his collaborator design a double free fall experiment to test equivalence
principle with free falling gyroscopes\cite{31g,31h}. There has also been
an interest in spin-gravitational coupling for a long
time\cite{31i,31j,31k,31l,21,32}.
\\

In a previous paper, the equation of motion of a spinning test particle
in gravitational field is deduced based on the coupling between the
spin of the particle and the gravitomagnetic field\cite{32}. In this
paper, we will study the post Newtonian approximation of that equation.
The post Newtonian approximation of the Papapetrou equation is
also studied. Then, we apply the equation to the problem of gyroscope
which is moving around the earth. Finally, qualitative results on
the motion of the gyroscope are given, and based on this calculation,
an experiment is proposed to  detect the effects of the violation of the
weak equivalence principle.\\

\section{Equation of Motion of a Spinning Particle}
\setcounter{equation}{0}

For the sake of integrity, we give a simple introduction to QGGR
and introduce some basic notations which is used in this paper. Details
on QGGR can be found in literatures \cite{08,09,10,11,12,13,14,16}. In gauge
theory of gravity, the most fundamental quantity is gravitational
gauge field $C_{\mu}(x)$, which is the gauge potential
corresponding to gravitational gauge symmetry. Gauge field
$C_{\mu}(x)$ is a vector in the corresponding Lie algebra, which
is called gravitational Lie
algebra. So $C_{\mu}(x) = C_{\mu}^{\alpha} (x)
\hat{P}_{\alpha} (\mu, \alpha = 0,1,2,3)$, where
$C_{\mu}^{\alpha}(x)$ is the component field and $\hat{P}_{\alpha}
= -i \frac{\partial}{\partial x^{\alpha}}$ is the  generator of
global gravitational gauge group. The gravitational gauge covariant
derivative is given by
\be \label{2.1}
D_{\mu} = \partial_{\mu} -
i g C_{\mu} (x) = G_{\mu}^{\alpha} \partial_{\alpha},
\ee
where $g$ is the gravitational coupling constant and matrix $G =
(G_{\mu}^{\alpha}) = ( \delta_{\mu}^{\alpha} - g C_{\mu}^{\alpha}
)$. Its inverse matrix is $G^{-1} = \frac{1}{I - gC} = (G^{-1
\mu}_{\alpha})$. Using matrix $G$ and $G^{-1}$, we can define two
important composite operators
\be \label{2.2}
g^{\alpha \beta} =
\eta^{\mu \nu} G^{\alpha}_{\mu} G^{\beta}_{\nu},
\ee
\be \label{2.3}
g_{\alpha \beta} = \eta_{\mu \nu} G_{\alpha}^{-1 \mu}
G_{\beta}^{-1 \nu},
\ee
which are widely used in QGGR. In QGGR,
space-time is always flat and space-time metric is always
Minkowski metric, so $g^{\alpha\beta}$ and $g_{\alpha\beta}$ are
no longer space-time metric. They are only two composite operators
which  consist of gravitational gauge field. The  field strength
of gravitational gauge field is defined by
\be \label{2.4}
F_{\mu\nu} (x) \define \frac{1}{-ig} \lbrack D_{\mu}~~,~~D_{\nu}
\rbrack = F_{\mu\nu}^{\alpha}(x) \cdot \hat{P}_{\alpha}
\ee
where
\be \label{2.5}
F_{\mu\nu}^{\alpha} = G_{\mu}^{\beta}
\partial_{\beta} C_{\nu}^{\alpha} -G_{\nu}^{\beta}
\partial_{\beta} C_{\mu}^{\alpha}.
\ee
\\

In QGGR, gravitational gauge field
$C_{\mu}^{\alpha}$ is a spin-2 tensor field. The field equation
of gravitational gauge field given by the least action principle
is equivalent to the Einstein's field equation in General
Relativity\cite{12,13,14}. \\

According to literature \cite{32}, the equation of motion of a spinning
test particle in gravitational field is
\be \label{2.6}
\frac{D p^{\gamma}}{D \tau}
= g^{\gamma \alpha}
\left \lbrack
\nabla_{\alpha} \left ( \frac{g}{2} g_{\beta \delta}
J^{\rho\sigma} F^{\delta}_{\rho \sigma}
\right )
- \nabla_{\beta} \left ( \frac{g}{2} g_{\alpha \delta}
J^{\rho\sigma} F^{\delta}_{\rho \sigma}
\right )
\right \rbrack \frac{{\rm d} x^{\beta}}{{\rm d} \tau},
\ee
where  $J^{\rho\sigma}$ is the spin tensor of the particle, and
\be \label{2.7}
\frac{D p^{\gamma}}{D \tau}
= \frac{{\rm d} p^{\gamma}}{{\rm d} \tau}
+ \Gamma^{\gamma}_{\alpha\beta}
p^{\alpha} \frac{{\rm d} x^{\beta}}{{\rm d} \tau},
\ee
or equivalently
\be \label{2.8}
\frac{D }{D \tau}
\left ( p^{\gamma}
+ \frac{g}{2} J^{\rho\sigma} F^{\gamma}_{\rho \sigma}
\right )
= \frac{g}{2} g^{\gamma \alpha} g_{\beta \delta}
\nabla_{\alpha} \left (
J^{\rho\sigma} F^{\delta}_{\rho \sigma}
\right )
 \frac{{\rm d} x^{\beta}}{{\rm d} \tau}.
\ee
Equation (\ref{2.6}) can be written into the following form
\be \label{2.9}
\frac{D p^{\alpha}}{D \tau}
= f_s^{ \alpha},
\ee
where $f_s^{\alpha}$ is the interaction force originated from the coupling
of spin and gravitomagnetic field,
\be \label{2.10}
f_s^{\alpha} = g^{\alpha\gamma} \left \lbrack
\partial_{\gamma} \left (
\frac{g}{2} g_{\beta \delta}
J^{\rho\sigma} F_{\rho\sigma}^{\delta} \right )
- \partial_{\beta} \left (
\frac{g}{2} g_{\gamma \delta}
J^{\rho\sigma} F_{\rho\sigma}^{\delta}
\right ) \right \rbrack
\frac{{\rm d} x^{\beta}}{{\rm d} \tau}.
\ee
\\

\section{Post Newtonian Approximation}
\setcounter{equation}{0}

Now, let's discuss its the post Newtonian approximation. The standard
procedure of the post Newtonian approximation can be found in text
book\cite{33,34}. In gauge theory of gravity, post Newtonian approximation
is done in a similar way. Let'e consider the case that a gyroscope moving
around the earth. Then its post Newtonian approximations give
\be \label{3.1}
\left \lbrace
\ba{rcl}
g_{00} & = & \eta_{00} - 2 \phi + o(\bar v^4) \\
&&\\
g_{ij} & = & \eta_{ij} - 2 \delta_{ij} \phi + o(\bar v^4)\\
&&\\
g_{i0 }&=& g_{0i } = \zeta_i + o(\bar v^5)\\
&&\\
U^0 &=& 1 + o(\bar v^2)\\
&&\\
U^i &=& v^i + o(\bar v^3)\\
&&\\
\frac{\partial}{\partial x^i} & \sim & \frac{1}{ \bar r} \\
&&\\
\frac{\partial}{\partial t} & \sim & \frac{\bar v}{\bar r}
\ea
\right . ,
\ee
where $\phi$ and $\zeta_i$ are given by the following relations
\be \label{3.2}
\phi = - \frac{G M_{\oplus}}{r},
\ee
\be \label{3.3}
\svec{\zeta} = \frac{2 G}{r^3}
(\svec{x} \times \svec{J_{\oplus}}).
\ee
In above relations, $M_{\oplus}$ and $\svec{J_{\oplus}}$ are mass and spin angular
momentum of the source. The leading terms of gravitational gauge fields are
\be \label{3.4}
\left \lbrace
\ba{l}
g C_i^j  =  g C_j^i = - \delta_{ij} \phi + o(\bar v^4) \\
\\
g C_0^0  =¡¡ \phi +  o(\bar v^4) \\
\\
g C_0^i  = -g C_i^0 = \frac{1}{2} \zeta_i + o(\bar v^5)
\ea
\right . .
\ee
The gravitational gauge field strength $F^{\alpha}_{\mu\nu}$ have the expansion
\be \label{3.5}
\left \lbrace
\ba{l}
g F_{ij}^k  =   \delta_{ik} (\partial_j \phi)
- \delta_{jk} (\partial_i \phi)  + o(\bar v^4/ \bar r) \\
\\
g F_{ij}^0  =¡¡- \frac{1}{2} ( \partial_i \zeta_j
- \partial_j \zeta_i) +  o(\bar v^5/ \bar r) \\
\\
g F_{i0}^k  = - g F_{0i}^k =
\frac{1}{2} \partial_i \zeta_k
+ \delta_{ik} \tdot{\phi} + o(\bar v^5/\bar r) \\
\\
g F_{i0}^0 = - g F_{0i}^0
= \partial_i \phi
+ o(\bar v^4/\bar r)
\ea
\right . .
\ee
If we use the following relations,
\be \label{3.6}
B_i^{\alpha} = - \frac{1}{2} \varepsilon_{ijk} F_{jk}^{\alpha},
\ee
\be \label{3.7}
E_i^{\alpha} =  F_{0i}^{\alpha},
\ee
we find that the gravitoelectromagnetic fields have the following approximations
\be \label{3.8}
\left \lbrace
\ba{rcl}
g B_i^k &=& -\frac{1}{2} \varepsilon_{ikm} (\partial_m \phi) + o(\bar v^4/ \bar r) \\
&&\\
g B_i^0 &=& \frac{1}{2} (\nabla \times \svec{\zeta})_i  + o(\bar v^5/ \bar r)\\
&&\\
g E_i^0 &=& - (\nabla \phi )_i + o(\bar v^4/ \bar r)\\
&&\\
g E_i^k &=& -(\frac{1}{2} \partial_i \zeta_k + \delta_{ik} \tdot{\phi}) + o(\bar v^5/ \bar r)
\ea
\right .
\ee
For spin tensor $J^{\mu\nu}$, we have
\be \label{3.9}
\left \lbrace
\ba{rcl}
J^{ij} & \sim & \bar J \\
&&\\
J^{i0} & \sim & \bar v \bar J
\ea
\right .
\ee
\\

The quantity $\Gamma_{\alpha\beta}^{\gamma}$ is defined by
\be \label{3.10}
\Gamma_{\alpha\beta}^{\gamma}
= \frac{1}{2} g^{\gamma\delta}
\left (
\frac{\partial g_{\alpha \delta}}{\partial x^{\beta}}
+ \frac{\partial g_{\beta \delta}}{\partial x^{\alpha}}
-\frac{\partial g_{\alpha \beta}}{\partial x^{\delta}}
\right ).
\ee
In post Newtonian approximations, its leading contributions are
\be \label{3.11}
\left \lbrace
\ba{rcl}
\Gamma_{00}^i &=& \frac{\partial \phi}{\partial x^i}
+ \partial_i (2 \phi^2 + \psi) + \tdot{\zeta}_i
+ o(\bar v^6/ \bar r) \\
&&\\
\Gamma_{0j}^i &=& \frac{1}{2} (\partial_j \zeta_i - \partial_i \zeta_j)
- \delta_{ij} \tdot{\phi}  + o(\bar v^5/ \bar r) \\
&&\\
\Gamma_{jk}^i &=& - \delta_{ij} \partial_k \phi
- \delta_{ik} \partial_j \phi + \delta_{jk} \partial_i \phi
+ o(\bar v^4/ \bar r) \\
&&\\
\Gamma_{00}^0 &=& \tdot{\phi} + \tdot{\psi}
+ \svec{\zeta} \cdot \nabla \phi
+ o(\bar v^7/ \bar r) \\
&&\\
\Gamma_{0i}^0 &=& \partial_i \phi
+ \partial_i \psi + o(\bar v^6/ \bar r) \\
&&\\
\Gamma_{ij}^0 &=& - \frac{1}{2}
(\partial_i \zeta_j + \partial_j \zeta_i)
- \delta_{ij} \tdot{\phi} + o(\bar v^5/ \bar r) \\
\ea
\right . ,
\ee
where $\psi$ is determined by the following equation
\be \label{3.12}
\nabla^2 \psi = \frac{\partial^2 \phi}{\partial t^2}
+ 4 \pi G \left (\stackrel{2~~}{T^{00}} + \stackrel{2~~}{T^{ii}} \right).
\ee
\\

Using all above approximations, we can calculate the leading contribution
of the force $f_s^{\alpha}$ from equation (\ref{2.10}), which is
\be \label{3.13}
f_s^i = \partial_i \left[ \frac{1}{2} \svec{J}
\cdot (\nabla \times \svec{\zeta})   \right ]
+ (\svec{J} \times \nabla)_i \left [ \partial_0 \phi
+ \svec{v} \cdot \nabla \phi   \right ]
+ 2 \svec{J} \cdot (\svec{v} \times \nabla) (\partial_i \phi),
\ee
\be \label{3.14}
f_s^0 = \svec{J} \cdot (\svec{v} \times \nabla)
\left [  - \partial_0 \phi + \svec{v} \cdot \nabla \phi \right ]
+ \frac{1}{2} (\svec{v} \cdot \nabla)
 \left [  \svec{J} \cdot ( \nabla  \times \svec{\zeta} )
 \right ].
\ee
We can see that $f_s^i$ is of order $\bar v^3 \bar J / \bar r^2$, and
$f_s^0$ is of order $\bar v^4 \bar J / \bar r^2$. \\

In general relativity, under pole-dipole approximation, the equation of
motion of a spinning particle is given by the following Papapetrou
equation
\be \label{3.16}
\frac{D}{D \tau} \left (
m u^{\alpha} + u_{\beta} \frac{D J^{\alpha\beta}}{D \tau}
\right )
- \frac{1}{2} J^{\beta \gamma} u^{\delta}
R^{\alpha}_{ \delta \beta \gamma }
 = 0,
\ee
where $u^{\alpha} = \frac{{\rm d} x^{\alpha}}{{\rm d} \tau}$
is the four-velocity, and $R^{\alpha}_{\delta\beta\gamma}$
is the curvature tensor. Now, let's discuss its post Newtonian
approximation. In most cases,
\be \label{3.17}
 \frac{D J^{\alpha\beta}}{D \tau}
\ll m,
\ee
so,
\be \label{3.18}
 u_{\beta} \frac{D J^{\alpha\beta}}{D \tau}
\ll m u^{\alpha}.
\ee
Therefore, in leading order approximation, the term
$u_{\beta} \frac{D J^{\alpha\beta}}{D \tau}$ in the Papapetrou
equation can be omitted. Then in post Newtonian approximation,
equation (\ref{3.16}) simplifies to
\be \label{3.19}
\frac{D p^{\alpha}}{D \tau} =
g  J^{\rho \sigma} u^{\beta}
\left ( \partial_{\beta} \partial_{\sigma} C_{\rho}^{\alpha}
- \eta^{\alpha\alpha_1} \eta_{\rho\beta_1}
\partial_{\alpha_1} \partial_{\sigma} C_{\beta}^{\beta_1}
\right ).
\ee
Compare it with equation (\ref{2.9}), we find that, in the present
case, the interaction force originated from the coupling between
the spin and gravitomagnetic field is
\be \label{3.20}
f_{s}^{\prime \alpha} =
g  J^{\rho \sigma} u^{\beta}
\left ( \partial_{\beta} \partial_{\sigma} C_{\rho}^{\alpha}
- \eta^{\alpha\alpha_1} \eta_{\rho\beta_1}
\partial_{\alpha_1} \partial_{\sigma} C_{\beta}^{\beta_1}
\right ).
\ee
After applying results  (\ref{3.1}) and (\ref{3.4}),
we can prove that
\be \label{3.21}
f_s^{\prime i} = \partial_i \left[ \frac{1}{2} \svec{J}
\cdot (\nabla \times \svec{\zeta})   \right ]
+ (\svec{J} \times \nabla)_i \left [ \partial_0 \phi
+ \svec{v} \cdot \nabla \phi   \right ]
+ 2 \svec{J} \cdot (\svec{v} \times \nabla) (\partial_i \phi),
\ee
\be \label{3.22}
f_s^{\prime 0} = \svec{J} \cdot (\svec{v} \times \nabla)
\left [  - \partial_0 \phi + \svec{v} \cdot \nabla \phi \right ]
+ \frac{1}{2} (\svec{v} \cdot \nabla)
 \left [  \svec{J} \cdot ( \nabla  \times \svec{\zeta} )
 \right ].
\ee
Compare them with equations (\ref{3.13}) and (\ref{3.14}), we find
that, in post Newtonian approximation, the force $f_s^{\prime \alpha}$
is the same as $f_s^{\alpha}$. So, the equation of motion (\ref{2.6})
and the Papapetrou equation give out the same results in the post
Newtonian approximation. \\

\section{Spin-Spin Interactions Force}
\setcounter{equation}{0}

Now, let's change the forms of equations (\ref{3.13}) and (\ref{3.14})
into more explicit forms. When we discuss the case that a gyroscope moves
around the earth, the post Newtonian
field $\phi$ and $\svec{\zeta}$ are given by equation (\ref{3.2}) and
(\ref{3.3}) respectively. Then in leading term approximations, equation
(\ref{3.13}) is changed into
\be \label{4.1}
\ba{rcl}
f_s^i &=& -3 G (\svec{J} \cdot \svec{J}_{\oplus}) \frac{r_i }{r^5}
-3 G (\svec{r} \cdot \svec{J}_{\oplus}) \frac{J_i }{r^5}
- 3 G (\svec{r} \cdot \svec{J}) \frac{J_{\oplus i}}{r^5}
+ 15 G (\svec{r} \cdot \svec{J}) (\svec{r} \cdot \svec{J}_{\oplus})
\frac{r_i}{r^7} \\
&&\\
&& + 3 G M_{\oplus} \frac{(\svec{J} \times \svec{v})_i}{r^3}
- 3 G M_{\oplus} (\svec{v} \cdot \svec{r})
  \frac{(\svec{J} \times \svec{r})_i}{r^5}
- 6 G M_{\oplus} [ \svec{J} \cdot (\svec{v} \times \svec{r}) ]
    \frac{r_i}{r^5}
+ (\svec{J} \times \nabla)_i \frac{\partial \phi}{\partial t},
\ea
\ee
and (\ref{3.14}) is changed into
\be \label{4.2}
\ba{rcl}
f_s^0 &=& -3 G (\svec{v} \cdot \svec{J}) (\svec{r} \cdot \svec{J}_{\oplus})
\frac{1}{r^5}
- 3 G (\svec{v} \cdot \svec{J}_{\oplus}) (\svec{r} \cdot \svec{J})
\frac{1}{r^5}
- 3 G (\svec{J} \cdot \svec{J}_{\oplus}) (\svec{r} \cdot \svec{v})
\frac{1}{r^5} \\
&&\\
&& + 15 G (\svec{r} \cdot \svec{J}) (\svec{r} \cdot \svec{J}_{\oplus})
    (\svec{r} \cdot \svec{v}) \frac{1}{r^7}
      - 3 G M_{\oplus} [ \svec{J} \cdot (\svec{v} \times \svec{r})]
    (\svec{r} \cdot \svec{v}) \frac{1}{r^5}
    - \svec{J} \cdot (\svec{v} \times \nabla) \frac{\partial \phi}{\partial t}.
\ea
\ee
\\

In equation (\ref{4.1}), the first four terms are from spin-spin interactions,
the next two terms are from spin-orbit interactions.
We can see that the spin-spin interaction force has the following form
\be \label{4.3}
\svec{f}_{ss} = -3 G (\svec{J} \cdot \svec{J}_{\oplus}) \frac{\svec{r} }{r^5}
-3 G (\svec{r} \cdot \svec{J}_{\oplus}) \frac{\svec{J} }{r^5}
- 3 G (\svec{r} \cdot \svec{J}) \frac{\svec{J}_{\oplus }}{r^5}
+ 15 G (\svec{r} \cdot \svec{J}) (\svec{r} \cdot \svec{J}_{\oplus})
\frac{\svec{r}}{r^7},
\ee
which is symmetric under the exchange of two spin operators. So, the spin-spin
interaction force is proportional to the spin of each particle, but decreases
as the inverse quadruplicate of the distances. Equation (\ref{4.3}) can
be changed into another form
\be \label{4.4}
f^i_{ss} = G \frac{J_j M^i_{jk} J_{\oplus k}  }{r^4},
\ee
where
\be \label{4.5}
M^i_{jk} = -3 \delta_{jk} \hat{r}_i
-3 \delta_{ij} \hat{r}_k
-3 \delta_{ik} \hat{r}_j
+ 15 \hat{r}_i \hat{r}_j \hat{r}_k.
\ee
In above equation, $\hat{r} = \frac{\svec{r}}{r}$ is a unit vector.
It can be seen that $M^i_{jk}$ is symmetric under exchange of any
two indexes of $i$, $j$ and $k$.
\\

\section{Test Gravity Theory With a Gyroscope}
\setcounter{equation}{0}

 Now, we discuss a classical test of gravity theory. Suppose that
 a gyroscope is placed in an orbit around the earth. For the earth,
 its gravitational field is quite stable, so the time derivative
 of the earth's gravitational potential almost vanishes
\be \label{5.1}
\frac{\partial \phi}{\partial t} = 0.
\ee
In the test, we select a special orbit so that the spin axis of the earth
is along the direction normal to the plane of the orbit. in this case
\be \label{5.2}
\svec{r} \cdot \svec{J}_{\oplus}
= \svec{v} \cdot \svec{J}_{\oplus}
= 0.
\ee
In this case, the time component and the space component of
the spin-dependent force $f_s^0$ and $f_s^i$   are respectively
simplified into
\be \label{5.3}
\ba{rcl}
f_s^i &=& -3 G (\svec{J} \cdot \svec{J}_{\oplus}) \frac{r_i }{r^5}
- 3 G (\svec{r} \cdot \svec{J}) \frac{J_{\oplus i}}{r^5}
+ 3 G M_{\oplus} \frac{(\svec{J} \times \svec{v})_i}{r^3}\\
&&\\
&&- 3 G M_{\oplus} (\svec{v} \cdot \svec{r})
  \frac{(\svec{J} \times \svec{r})_i}{r^5}
- 6 G M_{\oplus} [ \svec{J} \cdot (\svec{v} \times \svec{r}) ]
    \frac{r_i}{r^5},
\ea
\ee
\be \label{5.4}
f_s^0 =
- 3 G (\svec{J} \cdot \svec{J}_{\oplus}) (\svec{r} \cdot \svec{v})
\frac{1}{r^5}
      - 3 G M_{\oplus} [ \svec{J} \cdot (\svec{v} \times \svec{r})]
    (\svec{r} \cdot \svec{v}) \frac{1}{r^5}.
\ee
Then, we let the spin axis of the gyroscope be in the plane
of the orbit. So, the spin axis of the earth $\svec{J}_{\oplus}$
is perpendicular to the spin axis of the gyroscope $\svec{J}$,
\be \label{5.7}
\svec{J}_{\oplus} \cdot \svec{J} = 0,
\ee
and $\svec{r} \times \svec{v}$ is perpendicular to the orbit plane,
and therefore to the spin axis of the gyroscope
\be \label{5.7a}
\svec{J} \cdot (\svec{r} \times \svec{v}) = 0.
\ee
In this case, the time component of the spin-dependent force $f_s^0$
vanishes, and space-component of
 the spin-dependent force finally becomes
\be \label{5.8}
\svec{f}_s =
- 3 G (\svec{r} \cdot \svec{J}) \frac{\svec{J}_{\oplus }}{c^2 r^5}
+ 3 G M_{\oplus} \frac{(\svec{J} \times \svec{v})}{c^2 r^3}
- 3 G M_{\oplus} (\svec{v} \cdot \svec{r})
  \frac{(\svec{J} \times \svec{r})}{c^2 r^5}.
\ee
\\

In post newtonian approximation, $g_{\alpha\beta}$ is given by
(\ref{3.1}). In spherical coordinate system, we can prove that
the non-vanishing components of $g_{\alpha\beta}$ are
\be \label{5.8a1}
\left \lbrace
\ba{rcl}
g_{tt} & = & - B(r) \\
&&\\
g_{rr} & = & A(r) \\
&&\\
g_{\theta\theta} &=& A(r) r^2 \\
&&\\
g_{\varphi \varphi} &=& A(r) r^2 \sin^2 \theta \\
&&\\
g_{\varphi t} &=& g_{t \varphi} = C(r) \sin^2 \theta
\ea
\right . ,
\ee
where
\be \label{5.8a2}
\left \lbrace
\ba{rcl}
A(r) &=& 1 - 2 \phi = 1 + \frac{2 G M_{\oplus}}{r} \\
&&\\
B(r) &=& 1 + 2 \phi = 1 - \frac{2 G M_{\oplus}}{r} \\
&&\\
C(r) &=& - \frac{2 G J_{\oplus}}{c^4 r}
\ea
\right . .
\ee
The non-vanishing component of the inverse matrix of $g_{\alpha\beta}$
is
\be \label{5.8a3}
\left \lbrace
\ba{rcl}
g^{tt} & = & - \frac{1}{B(r)} \\
&&\\
g^{rr} & = & \frac{1}{A(r)} \\
&&\\
g^{\theta\theta} &=& \frac{1}{A(r) r^2} \\
&&\\
g^{\varphi \varphi} &=& \frac{1}{A(r) r^2 \sin^2 \theta} \\
&&\\
g^{\varphi t} &=& g^{t \varphi} = \frac{C(r)}{r^2}
\ea
\right . .
\ee
$\Gamma_{\alpha\beta}^{\gamma}$ is defined by (\ref{3.10}).
In spherical coordinate system, its non-vanishing components are
\be \label{5.8a4}
\left \lbrace
\ba{ll}
\Gamma^t_{tr} = \Gamma^t_{rt} = \frac{B'(r)}{2 B(r)}
& \Gamma^t_{r \varphi} = \Gamma^t_{\varphi r}
= \left [ - \frac{C'(r)}{2 B(r)} + \frac{A(r) C(r)}{r}
\right ] \sin^2 \theta \\
&\\
\Gamma^r_{tt} =  \frac{B'(r) }{2 A(r)}
 & \Gamma^t_{\theta\varphi} = \Gamma^t_{\varphi\theta}
= \left [- \frac{C(r) }{2 B(r)} + \frac{A(r) C(r)}{2}
\right] \sin 2\theta \\
&\\
\Gamma^r_{t \varphi} = \Gamma^r_{\varphi t} =
- \frac{C'(r)}{2 A(r)} \sin^2 \theta    ~~~~~~~
& \Gamma^r_{rr} = \frac{A'(r)}{2A(r)} \\
&\\
\Gamma^r_{\theta\theta} =
-\frac{r^2 A'(r)}{2 A(r)} - r
& \Gamma^r_{\varphi\varphi}
= \left [ -r - \frac{r^2 A'(r)}{2 A(r)}
\right ] \sin^2 \theta \\
&\\
\Gamma^{\theta}_{t \varphi} = \Gamma^{\theta}_{\varphi t}
= - \frac{C(r)}{ 2 r^2 A(r)} \sin 2 \theta
& \Gamma^{\theta}_{r \theta} = \Gamma^{\theta}_{\theta r}
= \frac{A'(r)}{2 A(r)} + \frac{1}{r} \\
&\\
\Gamma^{\theta}_{\varphi \varphi } = -\sin \theta \cos \theta
& \Gamma^{\varphi}_{t r} = \Gamma^{\varphi}_{r t}
= \frac{C'(r)}{2 r^2 A(r)}\\
&\\
\Gamma^{\varphi}_{t \theta} = \Gamma^{\varphi}_{\theta t}
= \frac{C(r)}{r^2 A(r)} {\rm ctg}  \theta
& \Gamma^{\varphi}_{r \varphi} = \Gamma^{\varphi}_{\varphi r}
= \frac{1}{r} + \frac{A'(r)}{2 A(r)} \\
&\\
\Gamma^{\varphi}_{\theta \varphi} = \Gamma^{\varphi}_{\varphi \theta}
= {\rm ctg} \theta
\ea
\right . .
\ee
\\

Now, let's discuss the equation of motion of the gyroscope, which
is given by equation (\ref{2.9}). Using above results
on $\Gamma^{\gamma}_{\alpha\beta}$, we can obtain
\be \label{5.9}
\ba{rl}
\frac{{\rm d}^2 r}{{\rm d} \tau^2} &
+ \frac{A'(r)}{2 A(r)} \left ( \frac{{\rm d}r}{{\rm d} \tau} \right )^2
+ \frac{B'(r)}{2 A(r)} \left (  \frac{{\rm d} t}{{\rm d} \tau} \right )^2
- \left [ r + \frac{r^2 A'(r)}{2A(r)} \right ]
 \left ( \frac{{\rm d} \theta}{{\rm d} \tau} \right )^2 \\
 &\\
& - \left [ r + \frac{r^2 A'(r) }{2A(r)} \right ]
\sin ^2 \theta \left ( \frac{{\rm d} \varphi}{{\rm d} \tau} \right )^2
- C'(r) \sin^2 \theta \frac{{\rm d} t}{{\rm d} \tau}
\frac{{\rm d} \varphi }{{\rm d} \tau}
= \frac{f^r_s}{m},
\ea
\ee
\be \label{5.10}
\frac{{\rm d}^2 \theta}{{\rm d} \tau^2}
- \frac{C(r)}{r^2} \sin 2\theta  \frac{{\rm d} t}{{\rm d} \tau}
\frac{{\rm d} \varphi }{{\rm d} \tau}
+ \left [ \frac{2}{r} + \frac{A'(r)}{A(r)} \right ]
\frac{{\rm d} r}{{\rm d} \tau} \frac{{\rm d} \theta}{{\rm d} \tau}
- \sin \theta \cos \theta \left ( \frac{{\rm d} \varphi}{{\rm d} \tau} \right )^2
= \frac{f^{\theta}_s}{m },
\ee
\be \label{5.11}
\frac{{\rm d}^2 \varphi}{{\rm d} \tau^2}
+ \frac{C'(r)}{r^2} \frac{{\rm d} t}{{\rm d} \tau}
\frac{{\rm d} r }{{\rm d} \tau}
+ \frac{2 C(r)}{r^2} {\rm ctg} \theta \frac{{\rm d} t}{{\rm d} \tau}
\frac{{\rm d} \theta }{{\rm d} \tau}
+ \left [ \frac{2}{r} + \frac{A'(r)}{A(r)} \right ]
\frac{{\rm d}r}{{\rm d} \tau} \frac{{\rm d} \varphi}{{\rm d} \tau}
+ 2 {\rm ctg} \theta
\frac{{\rm d} \theta}{{\rm d} \tau} \frac{{\rm d} \varphi}{{\rm d} \tau}
=\frac{f^{\varphi}_s}{m},
\ee
\be \label{5.12}
\frac{{\rm d}^2 t}{{\rm d} \tau^2}
+ \frac{B'(r)}{B(r)} \frac{{\rm d}t}{{\rm d} \tau}
    \frac{{\rm d} r}{{\rm d} \tau}
+ \frac{3 C(r)}{r} \sin^2 \theta \frac{{\rm d} r}{{\rm d} \tau}
\frac{{\rm d} \varphi }{{\rm d} \tau}
    =\frac{f^t_s}{m}.
\ee
The spherical coordinate system is specially selected so that the
polar angle $\theta$ of the plane of the orbit is $\frac{\pi}{2}$.
But the plane of the orbit can not always exactly be $\frac{\pi}{2}$, for
there always exists small perturbation to the motion of the gyroscope.
If  the small perturbation to the polar angle of the plane
of the orbit is denoted by $\delta \theta$,
then we have
\be \label{5.15}
\theta = \frac{\pi}{2} + \delta \theta,
\ee
where $\delta \theta$ is
a first order infinitesimal quantity. If we only keep first order
infinitesimal quantity,
then equations (\ref{5.9}), (\ref{5.10}), (\ref{5.11}) and (\ref{5.12})
are changed into
\be \label{5.16}
\frac{{\rm d}^2 r}{{\rm d} \tau^2}
+ \frac{A'(r)}{2 A(r)} \left ( \frac{{\rm d}r}{{\rm d} \tau} \right )^2
+ \frac{B'(r)}{2 A(r)} \left (  \frac{{\rm d} t}{{\rm d} \tau} \right )^2
- \frac{(r^2 A(r))'}{2A(r)} \left ( \frac{{\rm d} \varphi}{{\rm d} \tau} \right )^2
- \frac{C'(r)}{A(r)} \frac{{\rm d} \varphi}{{\rm d} \tau}
= \frac{f^r_s}{m},
\ee
\be \label{5.17}
\frac{{\rm d}^2 \delta \theta}{{\rm d} \tau^2}
+ \frac{2}{r} \frac{{\rm d} r}{{\rm d} \tau} \frac{{\rm d} \delta \theta}{{\rm d} \tau}
+ \delta \theta ~ \left ( \frac{{\rm d} \varphi}{{\rm d} \tau} \right )^2
= \frac{f^{\theta}_s}{m },
\ee
\be \label{5.18}
\frac{{\rm d}^2 \varphi}{{\rm d} \tau^2}
+ \left [ \frac{2}{r} + \frac{A'(r)}{A(r)} \right ]
\frac{{\rm d}r}{{\rm d} \tau} \frac{{\rm d} \varphi}{{\rm d} \tau}
+ \frac{\rm d}{{\rm d}\tau}
\left ( \frac{C(r)}{\sqrt{G M_{\oplus} L}} \right )
\frac{{\rm d} \varphi}{{\rm d} \tau}
=\frac{f^{\varphi}_s}{m},
\ee
\be \label{5.18a}
\frac{{\rm d}^2 t}{{\rm d} \tau^2}
+ \left [ \frac{B'(r)}{B(r)}
+ \frac{3 C(r) \sqrt{G M_{\oplus} L} }{r^3} \right ]
\frac{{\rm d}t}{{\rm d} \tau}
    \frac{{\rm d} r}{{\rm d} \tau}
    =\frac{f^t_s}{m}.
\ee
\\

In order to go any further, we need to know the explicit form the the spin
dependent force $\svec{f}_s$, which is given by equation (\ref{5.8}).
In the present coordinate system, the z axis is selected to be along
the direction of $\svec{J}_{\oplus}$, so
\be \label{5.19}
\svec{J}_{\oplus} = J_{\oplus} \hat{k},
\ee
\be \label{5.20}
\svec{J} \times \svec{v} =
\left [
e J \sqrt{\frac{G M_{\oplus}}{L}} \sin \varphi \sin (\varphi - \varphi_0)
+ J \frac{\sqrt{G M_{\oplus} L}}{r} \cos (\varphi - \varphi_0)
\right ] \hat{k},
\ee
\be \label{5.21}
\svec{r} \cdot \svec{J} = r J \cos (\varphi - \varphi_0),
\ee
where $\varphi_0$ is the intersection angle between the spin $\svec{J}$
and semimajor axis, $L$ is the semilatus rectum, and $e$ is the eccentricity.
Then the spin-dependent force has the following form
\be \label{5.21a}
f_s^t = 0,
\ee
\be \label{5.22}
\svec{f}_s = f_s \hat{k},
\ee
where
\be \label{5.23}
f_s = \left [
- \frac{3 G J J_{\oplus}}{c^2 r^4}
+ \frac{3G M_{\oplus} J \sqrt{G M_{\oplus} L}}{c^2 r^4}
 \right ]   \cos (\varphi - \varphi_0) .
\ee
The above expression for the force $\svec{f}_s$ is in the Descartes
coordinate system. We need to change its form into that in spherical
coordinate system. The components of the force $\svec{f}_s$ in spherical
coordinate system are
\be \label{5.24}
\left \lbrace
\ba{l}
f_s^r = 0, \\
\\
f_s^{\theta} = - \frac{f_s}{r}, \\
\\
f_s^{\varphi} = 0.
\ea
\right .
\ee
Then, equations (\ref{5.16}), (\ref{5.17}), (\ref{5.18})
and (\ref{5.18a}) are changed into
\be \label{5.25}
\frac{{\rm d}^2 r}{{\rm d} \tau^2}
+ \frac{A'(r)}{2 A(r)} \left ( \frac{{\rm d}r}{{\rm d} \tau} \right )^2
+ \frac{B'(r)}{2 A(r)} \left (  \frac{{\rm d} t}{{\rm d} \tau} \right )^2
- \frac{(r^2 A(r))'}{2A(r)} \left ( \frac{{\rm d} \varphi}{{\rm d} \tau} \right )^2
- \frac{C'(r)}{A(r)} \frac{{\rm d} \varphi}{{\rm d} \tau}
= 0,
\ee
\be \label{5.26}
\frac{{\rm d}^2 \delta \theta}{{\rm d} \tau^2}
+ \frac{2}{r} \frac{{\rm d} r}{{\rm d} \tau} \frac{{\rm d} \delta \theta}{{\rm d} \tau}
+ \delta \theta ~ \left ( \frac{{\rm d} \varphi}{{\rm d} \tau} \right )^2
= - \frac{f_s}{m r},
\ee
\be \label{5.27}
\frac{{\rm d}^2 \varphi}{{\rm d} \tau^2}
+ \left [ \frac{2}{r} + \frac{A'(r)}{A(r)} \right ]
\frac{{\rm d}r}{{\rm d} \tau} \frac{{\rm d} \varphi}{{\rm d} \tau}
+ \frac{\rm d}{{\rm d}\tau}
\left ( \frac{C(r)}{\sqrt{G M_{\oplus} L}} \right )
\frac{{\rm d} \varphi}{{\rm d} \tau}
=0,
\ee
\be \label{5.27a}
\frac{{\rm d}^2 t}{{\rm d} \tau^2}
+ \left [ \frac{B'(r)}{B(r)}
+ \frac{3 C(r) \sqrt{G M_{\oplus} L} }{r^3} \right ]
\frac{{\rm d}t}{{\rm d} \tau}
    \frac{{\rm d} r}{{\rm d} \tau}
    =0.
\ee
\\

Equations (\ref{5.27}), (\ref{5.27a}) and (\ref{5.25}) yield three constants of
motion, which are
\be \label{5.28}
r^2  A(r) \frac{{\rm d} \varphi}{{\rm d} \tau }
\cdot {\rm exp}\left [\frac{C(r)}{  \sqrt{G M_{\oplus} L}} \right ]
 = J_0~~~(constant),
\ee
\be \label{5.30}
B(r) \frac{{\rm d} t}{{\rm d} \tau} \cdot
{\rm exp} \left [ \frac{2 G J_{\oplus} \sqrt{G M_{\oplus} L}}{c^4 r^3}
\right ]
= 1.
\ee
\be \label{5.29}
A(r) \left ( \frac{{\rm d}r}{{\rm d} \tau} \right )^2
- \frac{1}{B(r)}
+ \frac{J_0^2}{r^2 A(r)} e^{-2 C(r) / \sqrt{G M_{\oplus} L}}
- \frac{4 G J_{\oplus} J^2_0}{3 c^4 r^3 \sqrt{G M_{\oplus} L}}
- \frac{4 G J_{\oplus} J_0}{3 c^4 r^3}
 = - E ~~~(constant),
\ee
\\

Multiply both sides of equation (\ref{5.26}) with $r^2$, we get
\be \label{5.31}
\frac{\rm d}{\rm d \tau}
\left (  r^2 \frac{\rm d \delta \theta}{\rm d \tau}   \right )
+ \frac{J_0^2}{r^2} \delta \theta
= - \frac{f_s}{m  } r.
\ee
Because
\be \label{5.32}
\frac{\rm d}{\rm d \tau}
= \frac{\rm d \varphi }{\rm d \tau} \frac{\rm d}{\rm d \varphi}
= \frac{J_0}{r^2} \frac{\rm d}{\rm d \varphi},
\ee
we can change (\ref{5.31}) into the following simpler form
\be \label{5.33}
\frac{{\rm d}^2 \delta \theta}{{\rm d} \varphi^2}
+  \delta \theta
= - \frac{f_s r^3}{J_0^2 m  }.
\ee
The above equation (\ref{5.33}) is the typical equation of motion of
a forced oscillator.
Applying (\ref{5.23}), we get
\be \label{5.34}
\frac{f_s r^3}{J_0^2 m} =  b \cos (\varphi - \varphi_0)
(1 + e \cos \varphi),
\ee
with $b$  a constant, which are defined by
\be \label{5.36}
b = \left ( \frac{3 (G M_{\oplus})^{3/2}}{c^2 J_0^2 L^{1/2}}
    - \frac{3 G J_{\oplus}}{c^2 J_0^2 L}
    \right )
    \left ( \frac{J}{m} \right ) .
\ee
\\

The general solution of oscillator equation (\ref{5.33}) is
\be \label{5.38}
\delta \theta (\varphi) = \alpha_1 \cos \varphi
    + \alpha_2 \sin \varphi
    + \alpha_3 (\varphi) \cos \varphi
    + \alpha_4 (\varphi) \sin \varphi,
\ee
where $\alpha_1$ and $\alpha_2$ are two constants, and $\alpha_3 (\varphi)$
and $\alpha_4 (\varphi)$ are determined by the following two equations
\be \label{5.39}
\frac{\rm d \alpha_3 (\varphi)}{ \rm d \varphi}
= \frac{f_s r^3 }{J_0^2 m} \sin \varphi,
\ee
\be \label{5.40}
\frac{\rm d \alpha_4 (\varphi)}{ \rm d \varphi}
= -  \frac{f_s r^3 }{J_0^2 m} \cos \varphi.
\ee
So, we have
\be \label{5.36a}
\ba{rcl}
\alpha_3(\varphi) & = &
\frac{b \sin \varphi_0}{2} \varphi
+ \frac{b  \cos \varphi_0}{2} \sin^2 \varphi
- \frac{b \sin \varphi_0}{2} \sin \varphi \cos \varphi \\
&&\\
&& + \frac{e b \cos \varphi_0}{3} (1 - \cos^3 \varphi)
+ \frac{e b \sin \varphi_0}{3} \sin^3 \varphi,
\ea
\ee
\be \label{5.36b}
\ba{rcl}
\alpha_4 (\varphi) & = &
- \frac{b \cos \varphi_0}{2} \varphi
- \frac{b  \cos \varphi_0}{2} \sin \varphi \cos \varphi
- \frac{b \sin \varphi_0}{2} \sin^2 \varphi  \\
&&\\
&& - e b \cos \varphi_0 (\sin \varphi - \frac{1}{3} \sin^3 \varphi)
- \frac{e b \sin \varphi_0}{3} ( 1- \cos^3 \varphi).
\ea
\ee
Therefore, the general solution of equation (\ref{5.33}) becomes
\be \label{5.36c}
\ba{rcl}
\delta \theta (\varphi) & = &
\alpha_1 \cos \varphi + \alpha_2 \sin \varphi
-\frac{b}{2} \varphi \sin (\varphi - \varphi_0)
-\frac{b \sin \varphi_0}{2} \sin \varphi \\
&&\\
&& + \frac{e b }{3}
[ - \frac{3}{2} \cos \varphi_0
+ \cos (\varphi + \varphi_0)
+ \frac{1}{2} \cos (2 \varphi - \varphi_0) ].
\ea
\ee
\\

The orbital period $T$ of the gyroscope is
\be \label{5.41}
T = 2 \pi \sqrt{\frac{a_G^3}{G M_{\oplus}}},
\ee
with $a_G$ the semimajor axis of the orbit of the gyroscope
\be \label{5.42}
a_G =  \frac{L}{1 - e^2} .
\ee
Suppose that at the starting point, $\varphi = 0$. Then at
an arbitrary time $t$, the value of $\varphi $ angle is about
\be \label{5.43}
\varphi (t) \simeq  \frac{2 \pi t}{T}.
\ee
When the time $t$ is many times larger than the orbital period $T$,
we will have the following approximate solutions for $\alpha_3$
and $\alpha_4$
\be \label{5.44}
\alpha_3 (t) \simeq
\frac{(G M_{\oplus})^{1/2}}{2 a_G^{3/2}} b t \sin \varphi_0,
\ee
\be \label{5.45}
\alpha_4 (t) \simeq
- \frac{(G M_{\oplus})^{1/2}}{2 a_G^{3/2}} b t \cos \varphi_0.
\ee
If at the starting point, we make the plane of the orbit exact at
the plane with $\theta = \frac{\pi}{2}$, which requires that when
$t$ is small, or equivalently $\varphi$ is small, $\delta \theta$
should almost vanish. Applying this initial condition to (\ref{5.38}),
we can obtain
\be \label{5.46}
\alpha_1 = \alpha_2 = 0.
\ee
Therefore, the solution (\ref{5.36c}) becomes
\be \label{5.47}
\delta \theta (t)
 = - \theta_m (t) \sin (\varphi -\varphi_0),
\ee
with
\be \label{5.48}
\theta_m (t) = \left [
\frac{3 G M_{\oplus}}{2 c^2 (L a_G)^{3/2}}
- \frac{3 J_{\oplus} (G M_{\oplus})^{1/2}}{2 c^2 M_{\oplus} L^2 a_G^{3/2}}
\right ]
\left (  \frac{J}{m}  \right ) t.
\ee
$\theta_m (t)$ is the amplitude of the oscillation, which is linearly
increased with time. \\

\section{Observable Effects}
\setcounter{equation}{0}

In the last chapter, we have discussed the motion of a gyroscope. When the
plane the orbit is perpendicular to the spin axis of the earth, and the
spin axis of the gyroscope lies in the orbit plane, the motion of the gyroscope
in the $\hat{e}_{\theta}$ direction obeys the forced oscillator equation
of motion. The resonant force $\svec{f}_s$ originates from the interaction
between the spin of the gyroscope and the gravitomagnetic field of the earth.
The solution of the equation is given by (\ref{5.47}). Now, our question
is that whether this forced oscillation is observable, or the amplitude
$\theta_m (t)$ is large enough to be detectable? And which information
can we obtained from this experiment? \\

Inside the satellite that carries the gyroscope, the quantity
$\theta_m (t)$ is not directly measurable, what  is directly
measurable inside the satellite is the relative
displacement of the gyroscope inside the satellite. The relative
displacement $\delta r_{\theta} (t)$ is given by radius $r$ times
angular displacement $\delta \theta (t)$, that is
\be \label{6.1}
\delta r_{\theta} (t) = r \cdot \delta \theta (t).
\ee
Therefore, we have
\be \label{6.2}
\delta r_{\theta} (t)
 = - w_m (t) \sin (\varphi -\varphi_0),
\ee
with
\be \label{6.3}
w_m (t) = r \cdot \left [
\frac{3 G M_{\oplus}}{2 c^2 (L a_G)^{3/2}}
- \frac{3 J_{\oplus} (G M_{\oplus})^{1/2}}{2 c^2 M_{\oplus} L^2 a_G^{3/2}}
\right ]
\left (  \frac{J}{m}  \right ) t.
\ee
In order that the oscillation can be observed, its amplitude $w_m$
should be large enough. Now, let's first estimate its magnitude. There
are two terms in the right hand of (\ref{6.3}), which represent two
separate contributions. They are denoted by
\be \label{6.4}
w_1 (t) =
\frac{3 G M_{\oplus} r}{2 c^2 (L a_G)^{3/2}}
\left (  \frac{J}{m}  \right ) t,
\ee
\be \label{6.5}
w_2 (t) =
- \frac{3 r J_{\oplus} (G M_{\oplus})^{1/2}}{2 c^2 M_{\oplus} L^2 a_G^{3/2}}
\left (  \frac{J}{m}  \right ) t,
\ee
where $w_1$ represents the  spin-orbit interactions, and $w_2$
represents the spin-spin interactions. So, we have
\be \label{6.6}
w_m (t) = w_1 (t) + w_2 (t).
\ee
Suppose that the orbit is near the surface of the earth, in this case
\be \label{6.7}
L \simeq a_G \simeq r \simeq R_{\oplus} \simeq 6.38 \times 10^6 m.
\ee
It is known that the spin angular momentum of the earth is about
\be \label{6.8}
J_{\oplus} \simeq 5.92 \times 10^{33} ~~ kg \cdot m^2 /s.
\ee
Using all these approximations, we have the following estimations
\be \label{6.9}
w_1 (t) = 5.14 \times 10^{-9}
\left (  \frac{J}{m}  \right ) ~m/year,
\ee
\be \label{6.10}
w_2 (t) = -1.01 \times 10^{-10}
\left (  \frac{J}{m}  \right ) ~m/year,
\ee
where the unit of $J/m$ is in $m^2/s$. If $J/m$ is about $10^3 m^2/s$,
then
\be \label{6.11}
w_1 (t) = 5.14  ~\mu m/year,
\ee
\be \label{6.12}
w_2 (t) = -0.101  ~\mu m/year,
\ee
Both of them are large enough to be detectable. According to (\ref{6.6}),
we have
\be \label{6.13}
w_m (t) = 5.04  ~\mu m/year.
\ee
So, if the magnitude of $J/m$ is about $10^3 m^2/s$,
the oscillation amplitude of the gyroscope in the $\hat{e}_{\theta}$
direction increases about 5.04 micrometer per year, which is detectable. \\

In the above discussions,
it is found that, the dominant contribution to the increasing of the
oscillation amplitude comes from the spin-orbit contributions, which is about
51 times larger than that of the spin-spin interactions. Now, the question
is that, whether can we simultaneous detect the magnitude of both $w_1$
and $w_2$? The answer is that, if we want to detect both of them, we
need at least two gyroscopes. Suppose that we have another gyroscope,
which is in another satellite that is running around the earth in the
opposite direction but with the same orbit parameters.
In this case, the direction of the spin axis of the earth inverses, so
the oscillation of the gyroscope in the $\hat{e}_{\theta}$ direction
is
\be \label{6.14}
\delta r_{\theta} (t)
 = - w'_m (t) \sin (\varphi -\varphi_0),
\ee
where
\be \label{6.15}
w'_m (t) = w_1 (t) - w_2 (t).
\ee
After we detect both $w_m (t)$ and $w'_m (t)$, we can determine both
spin-orbit contribution $w_1 (t)$ and spin-spin contribution $w_2 (t)$
\be \label{6.16}
w_1 (t) = \frac{w_m (t) + w'_m (t)}{2},
\ee
\be \label{6.16}
w_2 (t) = \frac{w_m (t) - w'_m (t)}{2}.
\ee
\\

\section{Summary and Discussions}
\setcounter{equation}{0}

In this paper, based on the equation of motion of a spinning particle
in gravitational field, spin-dependent gravitational force is discussed.
This kind of gravitational force contains two part, one is the
spin-orbit interactions, and another is spin-spin interactions. Both
of them are spin-dependent force. This force originates from the coupling
between the spin of the particle and gravitoelectromagnetic field.
So, whether the results in this paper are reliable depends on whether
the coupling between spin and gravitoelectromagnetic field is correct.
It is known that, another important effects of gravity, the precession
of a gyroscope in gravitational field, also originate from the
spin-gravitomagnetic coupling\cite{33,34,18}. That is, the influence of the
spin-gravitomagnetic coupling to the motion of the spin is the precession
of a gyroscope, while its influence to the motion of particle is what
we discussed in this paper. It is know that the effect on the precession
of a gyroscope had already been detected by experiments recently\cite{35,36},
and experimental results are quite consistent with theoretical expectations,
which tells us that the spin-gravitomagnetic coupling does exist and is reliable.
As long as the spin-gravitomagnetic coupling has influences to the motion
of the  spin, it must have influence to the motion of the particle. \\

The spin-dependent force is expressed by equations (\ref{3.13})
and (\ref{3.14}). The existence of this force explicitly showed that
not all gravitational actions are inertial actions, which means
the violation of weak equivalence principle\cite{37}. We can consider
this violation from a traditional way. We may consider that
the satellite which carries the gyroscope is a free falling frame
of reference. According to the weak equivalence principle, no objects
in the free falling frame of reference can fell gravitational force,
and therefore all objects in that frame should be relatively static
or uniform rectilinear moving. But now, we know that, owing to the
spin dependent force, the gyroscope is neither relatively static
nor uniform rectilinear moving, but oscillating with increasing
amplitude. The existence of this oscillation means that the
gyroscope feels gravitational force in the free falling frame
of reference, which violates the weak equivalence principle.\\

For a long time, it is generally believed that, even if there
exist spin-spin interaction, it must be too weak to be detectable.
Indeed, the spin-spin or spin-orbit interaction force is extremely weak.
According to equation (\ref{5.8}), the spin-spin interaction
force is about
\be \label{7.1}
-  \frac{3 G J J_{\oplus}}{c^2 r^4}.
\ee
If we suppose that the angular momentum of the gyroscope is
about $10^4 kg m^2/s$, then this force is about
$8 \times 10^{-17} N$, which is indeed too weak to be directly
detectable, and is many orders smaller than the upper limit given
by experiments\cite{31a,31b,31c,31d,31e}.
But, this force will cause a resonance oscillation
in the $\hat{e}_{\theta}$ direction, which is detectable. Direct
measurement of this oscillation will be of great importance, for
it is the first direct observation of the violation of the
weak equivalence principle.
There is a long standing dispute on the reliability of the weak
equivalence principle. Some treated it as a corner stone of
general relativity, while some others criticized that it is not
strictly hold\cite{37,38,39}. Now, the experiment can serve as
a direct experimental solution of the dispute. \\

Because QGGR can return to general relativity in classical
level, and the corner stone of QGGR is gauge principle, our
conclusion is that the equivalence principle is not the
necessary starting point of general relativity. Therefore,
the violation of weak equivalence principle does not affect the
correctness of general relativity. On the contrary, if equivalence
principle is still considered as corner stone of general relativity,
the theory will be logically self-contradictory, for in that case,
the violation of weak equivalence principle is a result of the
equivalence principle. (Because Papapetrou equation is a direct result
of Einstein's field equation.) \\

\end{document}